\documentclass[aps,prl,showpacs,twocolumn,a4paper]{revtex4}
\usepackage{graphicx}
\usepackage{amsmath}
\usepackage{amsfonts}

\begin{document}

\title{Fractional Quantum Hall Effect of Hard-Core Bosons in Topological Flat Bands}
\author{Yi-Fei Wang$^{1,2}$, Zheng-Cheng Gu$^{3}$, Chang-De Gong$^{1,4}$, and D. N. Sheng$^{2}$}
\affiliation{$^1$Center for Statistical and Theoretical Condensed
Matter Physics, and Department of Physics, Zhejiang Normal
University, Jinhua 321004, China \\$^2$Department of Physics and
Astronomy, California State University, Northridge, California
91330, USA \\$^3$Kavli Institute for Theoretical Physics, University
of California, Santa Barbara, California 93106, USA
\\$^4$National Laboratory of Solid State Microstructures
and Department of Physics, Nanjing University, Nanjing 210093,
China}
\date{\today}

\begin{abstract}
Recent proposals of topological flat band (TFB) models have provided
a new route to realize the fractional quantum Hall effect (FQHE)
without Landau levels. We study hard-core bosons with short-range
interactions in two representative TFB models, one of which is the
well known Haldane model (but with different parameters). We
demonstrate that FQHE states emerge with signatures of even number
of quasi-degenerate ground states on a torus and a robust spectrum
gap separating these states from higher energy spectrum. We also
establish quantum phase diagrams for the filling factor $1/2$ and
illustrate quantum phase transitions to other competing
symmetry-breaking phases.
\end{abstract}

 \maketitle

{\it Introduction.---}The fractional quantum Hall effect (FQHE), one
of the most fascinating discoveries in two-dimensional (2D) electron
gas, has set up a paradigm to explore new topological phases in
other strongly correlated systems. As commonly believed, the FQHE
requires two basic ingredients: single-particle states with
nontrivial topology, and quenching of the kinetic energy compared to
interaction energy scale. However, despite of the seemingly
universal theoretical concepts, the FQHE has only been found in 2D
systems under a strong perpendicular magnetic field, i.e., in which
particles move in Landau levels (LLs). In rotating Bose-Einstein
condensate~\cite{Cooper} and optical lattice
systems~\cite{Demler,Palmer}, researchers have been interested in
generating an artificial uniform magnetic field, thus the bosonic
FQHE states are expected, but still due to the existence of LLs.

Haldane's honeycomb lattice model~\cite{Haldane} and other similar
lattice models~\cite{Yakovenko,Nagaosa} have two nontrivial
topological bands characterized by $\pm1$ Chern
numbers~\cite{Thouless, Niu}, demonstrating the integer quantum Hall
effect without LLs. However, these single-particle bands are still
highly dispersive, thus it is unlikely to realize the FQHE in such
systems. Recently, proposals of topological flat bands
(TFB)~\cite{Wen,Santos,Sun} shed new lights to overcome this
long-standing and hard problem. These TFB models belong to the same
topological class as the Haldane model and are distinct from other
flat bands with a zero Chern number~\cite{Wu}. The energy dispersion
in such a TFB is substantially reduced by tuning short-range hopping
parameters. In particular, based on the mechanism of quadratic band
touching ~\cite{Wen,Sun}, a series of TFB models have been
explicitly constructed with a flatness ratio (the ratio of the band
gap over bandwidth) reaching a high value between $20\sim50$.

Here, we focus on the possible bosonic FQHE in TFB models filled
with interacting hard-core bosons, since the TFB will be more likely
realized in optical lattices by manipulating bosonic cold
atoms~\cite{CWu1,Xing2,Stanescu}. Although TFB models possess both
ingredients to realize the FQHE, quantum phases in such systems are
determined by some competing effects, different from a LL problem.
The main effects are: i) the lattice effect and the residual kinetic
energy since a TFB is not strictly flat; ii) the Berry curvature of
a TFB has substantial momentum dependence representing a non-uniform
magnetic field effect in momentum space; iii) lattice symmetry
breaking may lead to other conventional ordered states for hard-core
bosons. Thus it is highly desirable to pursue a systematic numerical
study of such interacting TFB models.

In this letter, we present the exact diagonalization (ED)
calculations of two representative TFB models with the
nearest-neighbor (NN) and the next-nearest-neighbor (NNN) repulsions
$V_1$ and $V_2$. We find convincing numerical evidences of both the
$1/2$ and the $1/4$ bosonic FQHE phases which are characterized by
the formation of quasi-degenerate ground-state manifold (GSM) with
even number of states. The GSM carries a unit total Chern
number~\cite{Sheng}, which is a robust property of the system
protected by a finite energy spectrum gap. We also determine phase
diagrams for our systems and illustrate the quantum phase
transitions based on the calculations of the density structure
factors and the fidelity~\cite{SJGu} of the ground state (GS)
wavefunction.

\begin{figure}[!htb]
  \vspace{0.05in}
  \begin{minipage}[c]{0.5\textwidth}
  \includegraphics[scale=0.28]{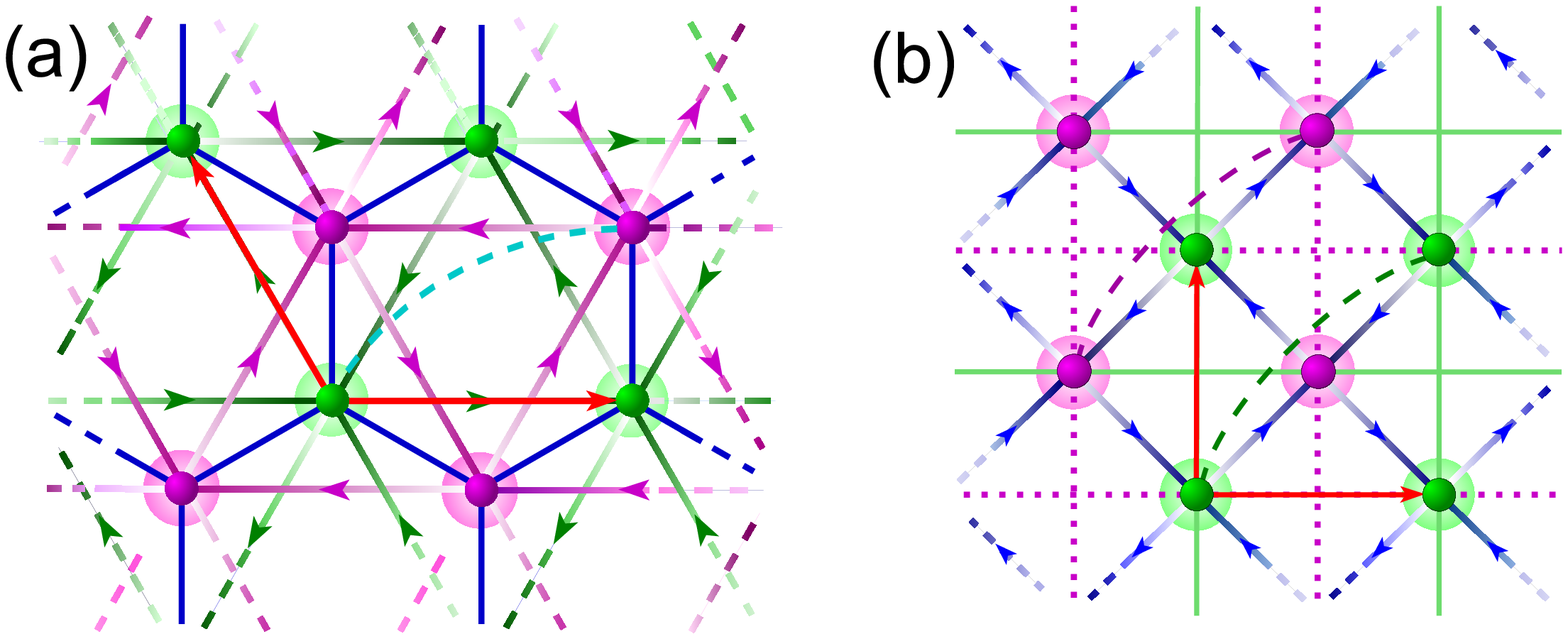}
\par\vspace{0pt}
\end{minipage}
  \vspace{-0.20in}
  \begin{minipage}[c]{0.5\textwidth}
  \includegraphics[scale=0.8]{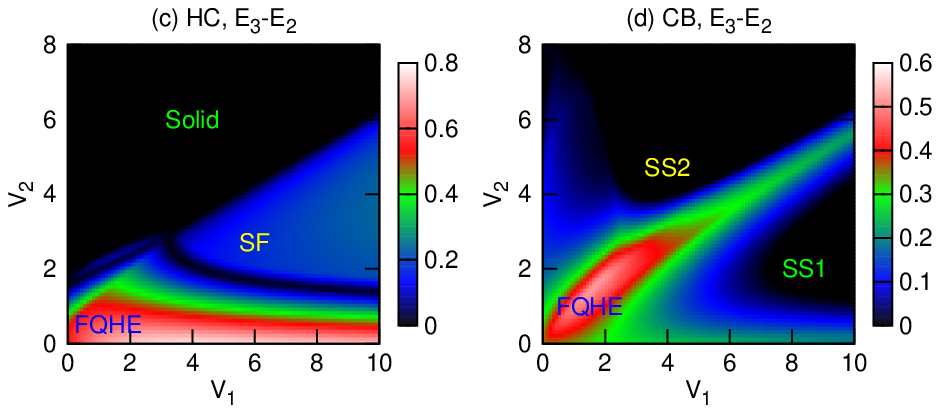}
\par\vspace{0pt}
\end{minipage}
  \vspace{-0.05in}
  \caption{(color online). (a) The Haldane model on
the honeycomb lattice and (b) the checkerboard model. The arrow
directions present the signs of the phases $\pm\phi$ in the NNN or
NN hopping terms. For the checkerboard lattice, the NNN hopping
amplitudes are $t^{\prime}$ ($-t^{\prime}$) along the solid (dotted)
lines. The NNNN hoppings are represented by the dashed curves in
both models. (c)-(d) Intensity plots of spectrum gaps in the
$V_1$-$V_2$ phase space at $\nu=1/2$ for (c) $24$-site honeycomb
lattice; (d) $24$-site checkerboard lattice. FQHE, SF, SS1/SS2, and
Solid label  estimated phase regions inferred from the spectrum-gap
plots and other information (see text).} \label{f.1}
\end{figure}

{\it Model Hamiltonians.---}The first model is the Haldane
model~\cite{Haldane} on the honeycomb (HC) lattice filled with
interacting hard-core bosons:
\begin{eqnarray}
H_{\rm HC}&=& -t^{\prime}\sum_{\langle\langle\mathbf{r}\mathbf{r}^{
\prime}\rangle\rangle}
\left[b^{\dagger}_{\mathbf{r}^{ \prime}}b_{\mathbf{r}}\exp\left(i\phi_{\mathbf{r}^{ \prime}\mathbf{r}}\right)+\textrm{H.c.}\right]\nonumber\\
&-&t\sum_{\langle\mathbf{r}\mathbf{r}^{ \prime}\rangle}
\left[b^{\dagger}_{\mathbf{r}^{\prime}}b_{\mathbf{r}}+\textrm{H.c.}\right]
-t^{\prime\prime}\sum_{\langle\langle\langle\mathbf{r}\mathbf{r}^{
\prime}\rangle\rangle\rangle}
\left[b^{\dagger}_{\mathbf{r}^{\prime}}b_{\mathbf{r}}+\textrm{H.c.}\right]\nonumber\\
&+&V_1\sum_{\langle\mathbf{r}\mathbf{r}^{ \prime}\rangle}
n_{\mathbf{r}}n_{\mathbf{r}^{\prime}}
+V_2\sum_{\langle\langle\mathbf{r}\mathbf{r}^{
\prime}\rangle\rangle}n_{\mathbf{r}}n_{\mathbf{r}^{\prime}}\label{e.1}
\end{eqnarray}
where $b^{\dagger}_{\mathbf{r}}$ creates a hard-core boson at site
$\mathbf{r}$, $n_{\mathbf{r}}$ is the boson number operator,
$\langle\dots\rangle$, $\langle\langle\dots\rangle\rangle$ and
$\langle\langle\langle\dots\rangle\rangle\rangle$ denote the NN, the
NNN and the next-next-nearest-neighbor (NNNN) pairs of sites,
respectively [Fig.~\ref{f.1}(a)]. We call this model as a
Haldane-Bose-Hubbard (HBH) model.

The second model is a variant version of the HBH model on the 2D
checkerboard (CB) lattice~\cite{Yakovenko,Xing1,Yao,Sun}:
\begin{eqnarray}
H_{\rm CB}&=& -t\sum_{\langle\mathbf{r}\mathbf{r}^{ \prime}\rangle}
\left[b^{\dagger}_{\mathbf{r}^{ \prime}}b_{\mathbf{r}}\exp\left(i\phi_{\mathbf{r}^{ \prime}\mathbf{r}}\right)+\textrm{H.c.}\right]\nonumber\\
&\pm&t^{\prime}\sum_{\langle\langle\mathbf{r}\mathbf{r}^{
\prime}\rangle\rangle}
\left[b^{\dagger}_{\mathbf{r}^{\prime}}b_{\mathbf{r}}+\textrm{H.c.}\right]
-t^{\prime\prime}\sum_{\langle\langle\langle\mathbf{r}\mathbf{r}^{
\prime}\rangle\rangle\rangle}
\left[b^{\dagger}_{\mathbf{r}^{\prime}}b_{\mathbf{r}}+\textrm{H.c.}\right]\nonumber\\
&+&V_1\sum_{\langle\mathbf{r}\mathbf{r}^{ \prime}\rangle}
n_{\mathbf{r}}n_{\mathbf{r}^{\prime}}
+V_2\sum_{\langle\langle\mathbf{r}\mathbf{r}^{
\prime}\rangle\rangle}n_{\mathbf{r}}n_{\mathbf{r}^{\prime}}\label{e.2}
\end{eqnarray}

{\it Topological flat bands.---}On the honeycomb lattice, if we
restrict the model with only NN and NNN hoppings, the best TFB has a
flatness ratio $7$~\cite{Santos}. By allowing the NNNN hoppings, we
numerically found several flatter bands with nonzero Chern numbers,
e.g. a flatness ratio about $50$ for the set of parameters, which
will be used here: $t=1$, $t^{\prime}=0.60$,
$t^{\prime\prime}=-0.58$ and $\phi=0.4\pi$. Using these parameters,
the TFB is gaped from quadratic touching~\cite{Yao} by breaking the
time reversal symmetry but preserving other lattice
symmetries~\cite{Sun}.

On the checkerboard lattice, we adopt the parameters of
Ref.~\cite{Sun} with an additional minus sign (to make the TFB as the
lower energy band), $t=-1$, $t^{\prime}=1/(2+\sqrt{2})$,
$t^{\prime\prime}=-1/(2+2\sqrt{2})$ and $\phi=\pi/4$, which leads to
a TFB with the flatness ratio about $30$.

{\it The $\nu=1/2$ phase diagrams.---} In the ED study, we consider
a finite system of $N_1\times N_2$ unit cells (total number of sites
$N_s=2\times N_1\times N_2$) with periodic boundary conditions
(PBC), denoting the number of bosons as $N_b$, and the filling
factor of the TFB is thus $\nu=N_b/(N_1N_2)$. In both models, the
amplitude of NN hopping $|t|$ is set as the unit of energy. We first
glance at the spectrum gaps of the two $24$-site ($2\times4\times3$)
lattices at the filling $\nu=1/2$ as shown in Fig.~\ref{f.1}(c) and
\ref{f.1}(d). $E_1$, $E_2$ and $E_3$ denote the energies of the
three lowest eigenstates. For the $\nu=1/2$ FQHE phase at the left
bottom corners in the $V_1$-$V_2$ space, there is a GSM with two
quasi-degenerate lowest eigenstates, and the GSM is separated from
higher eigenstates by a finite spectrum gap $E_3-E_2$ ($\gg
E_2-E_1$). The other rough phase regions for the possible superfluid
(SF), the supersolids (SS1/SS2) and the solid will be discussed
later. We have also obtained numerical results from larger lattice
sizes of $32$ ($2\times4\times4$), $36$ ($2\times6\times3$) and $40$
($2\times4\times5$) sites, and have confirmed both phase diagrams
are qualitatively correct.

\begin{figure}[!htb]
  \vspace{0.0in}
  \hspace{-0.08in}
  \includegraphics[scale=0.55]{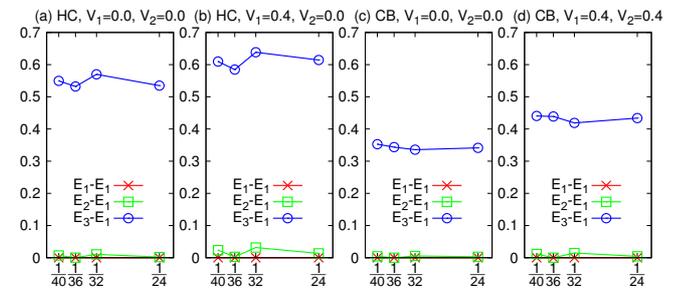}
  \vspace{-0.06in}
  \caption{(color online). $1/2$-FQHE spectrum gaps versus
  $1/N_s$: (a)-(b) honeycomb lattice; (c)-(d) checkerboard lattice.
  } \label{f.2}
\end{figure}

{\it Low energy spectrum and robust spectrum gap.---}We denote the
momentum vector $\mathbf{q}=(2\pi k_1/N_1,2\pi k_2/N_2)$ with
$(k_1,k_2)$ as integer quantum numbers. The GSM is defined as a set
of lowest states separated from other excited states by a finite
spectrum gap. If $(k_1,k_2)$ is the momentum sector for one of the
state in the GSM, we find that the other state should be obtained in
the sector $(k_1+N_b,k_2+N_b)$ [module $(N_1,N_2)$] . For $N_s=24$,
$36$ and $40$, the two states within the GSM of a $\nu=1/2$ FQHE
phase are indeed in different momentum sectors: $(0,0)$ and $(2,0)$
for $N_s=24$ and $N_s=40$, while $(0,0)$ and $(3,0)$ for $N_s=36$.
For $N_s=32$, both $N_b/N_1$ and $N_b/N_2$ are integers, thus both
states within the GSM are in the $(0,0)$ sector.

Now we check whether the spectrum gap $E_3-E_2$ remains in the
thermodynamic limit. As shown in Fig.~\ref{f.2}, when $N_s$
increases, the spectrum gap $E_3-E_2$ does not decrease, which
extrapolates to a finite value at large $N_s$ limit. Interestingly,
the spectrum gap $E_3-E_2$ is already quite large ($E_3-E_2\gg
E_2-E_1$) for hard-core boson system without additional interactions
($V_1$=$V_2$=0) [in Figs.~\ref{f.2}(a) and ~\ref{f.2}(c)]
demonstrating the robust $1/2$ FQHE. The spectrum gap can be
slightly enhanced with  a small $V_1$ and/or $V_2$
[Figs.~\ref{f.2}(b) and \ref{f.2}(d)].

By comparing the spectrum gap $E_3-E_2$ between two lattices for the
$V_1=V_2=0.0$ cases in Figs.~\ref{f.2}(a) and \ref{f.2}(c), we
notice that the gap $E_3-E_2$ in the honeycomb lattice is obviously
larger, which might be due to its larger flatness ratio. After
studies on a few more cases with smaller flatness ratios
$7$~\cite{Santos} and $30$ (with $t^{\prime}=0.40$,
$t^{\prime\prime}=-0.33$ and $\phi=0.5\pi$) on the honeycomb
lattice, we conclude that the flatter the TFB is, the larger the
spectrum gap $E_3-E_2$ can be, indicating a more robust FQHE,
although the global structure of the phase diagram does not change
much.

\begin{figure}[!htb]
  \vspace{0.05in}

\hspace{-0.03\textwidth}%
\begin{minipage}[c]{0.32\textwidth}
  \includegraphics[scale=0.36]{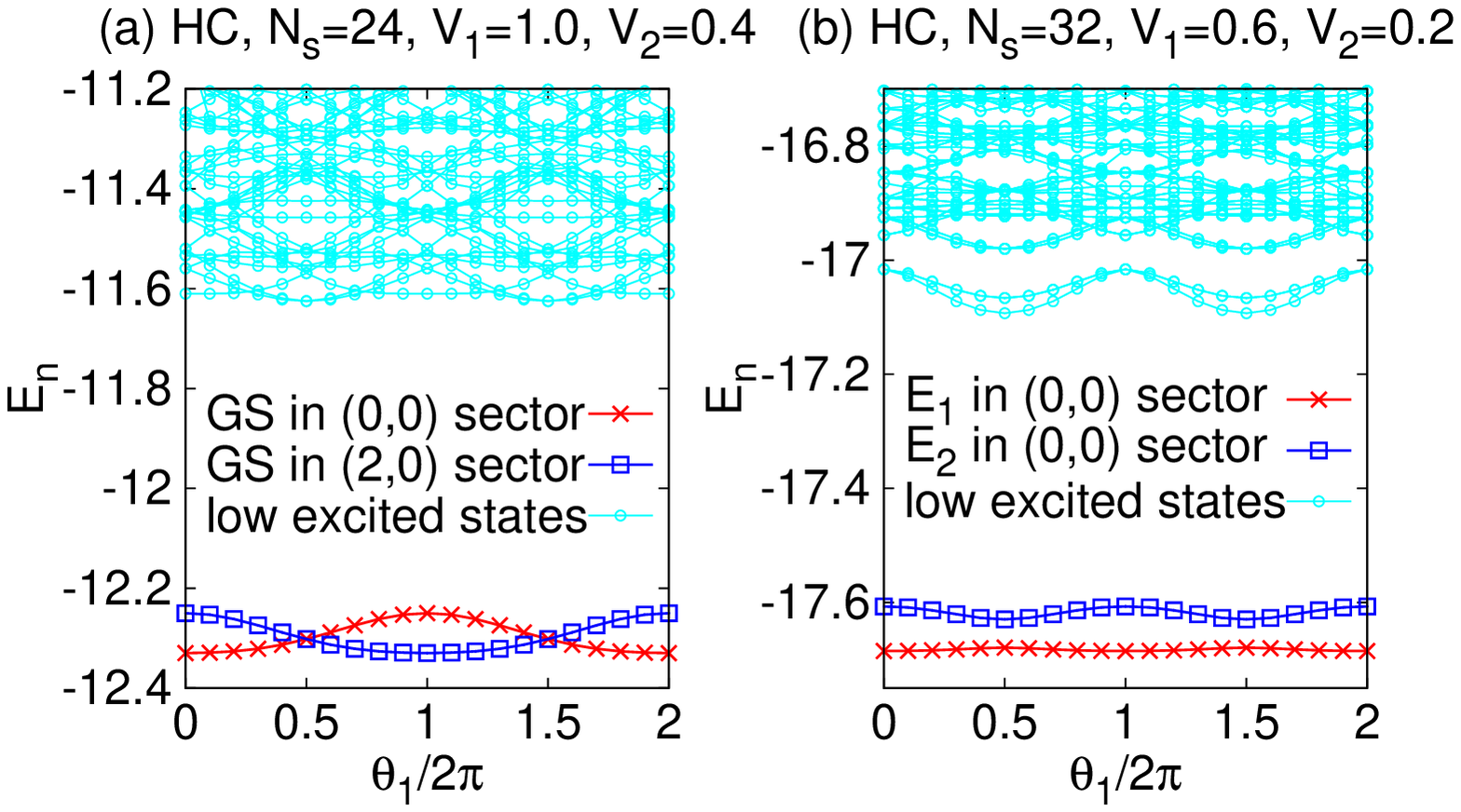}
\par\vspace{0pt}
\end{minipage}
\hspace{-0.004\textwidth}%
\begin{minipage}[c]{0.14\textwidth}
  \includegraphics[scale=0.53]{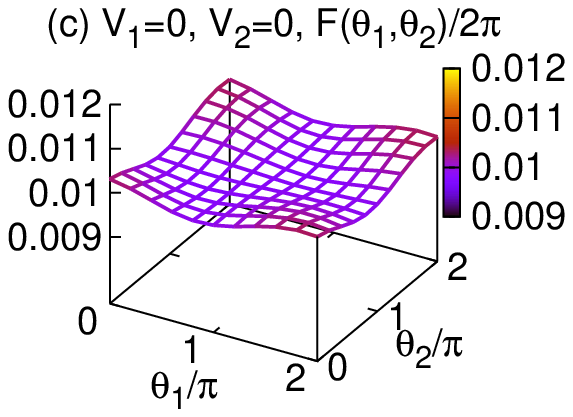}
\par\vspace{0pt}
\end{minipage}%
\vspace{-0.02in}
  \caption{(color online). (a)-(b) Low energy spectra versus
  $\theta_1$ at a fixed $\theta_2=0$
  for the honeycomb lattices at $\nu=1/2$.
   (c) $F(\theta_1,\theta_2)$ of a GSM of the 32-site honeycomb lattice. } \label{f.3}
\end{figure}

{\it Berry curvature and Chern number.---}Introducing two boundary
phases $\theta_1$ and $\theta_2$ as the generalized boundary
conditions in both directions, the Chern number~\cite{Thouless} (is
also the Berry phase in units of $2\pi$) of a many-body state is
given by an integral in the boundary phase space~\cite{Niu,Sheng}:
$C={{1}\over{2\pi}}\int\int d\theta_1\theta_2 F(\theta_1,\theta_2)$,
where the Berry curvature is given by $F(\theta_1,\theta_2)=\rm{Im}
\left(\left\langle {{\partial
\Psi}\over{\partial\theta_2}}\Big{|}{{\partial
\Psi}\over{\partial\theta_1}}\right\rangle -\left\langle {{\partial
\Psi}\over{\partial\theta_1}}\Big{|}{{\partial
\Psi}\over{\partial\theta_2}}\right\rangle\right)$. For each GSM of
$1/2$-FQHE phase with $N_s=24,~36,~40$, the two states are found to
evolve into each other with level crossing when tuning the boundary
phases [Fig.~\ref{f.3}(a)]. While for $N_s=32$,  with both states of
the GSM in the $(0,0)$ sector, each state evolves into itself when
tuning the boundary phases, and avoided level crossings appear
instead [Fig.~\ref{f.3}(b)] due to nonzero coupling between same
momentum states.  The GSM in the FQHE phase also has rather
smooth Berry curvature [Fig.~\ref{f.3}(c)] and shares a total Chern
number $C=1$.

\begin{figure}[!htb]
  \vspace{-0.15in}
  \hspace{0.0in}
\begin{minipage}[b]{0.5\textwidth}
\centering
\includegraphics[scale=0.55]{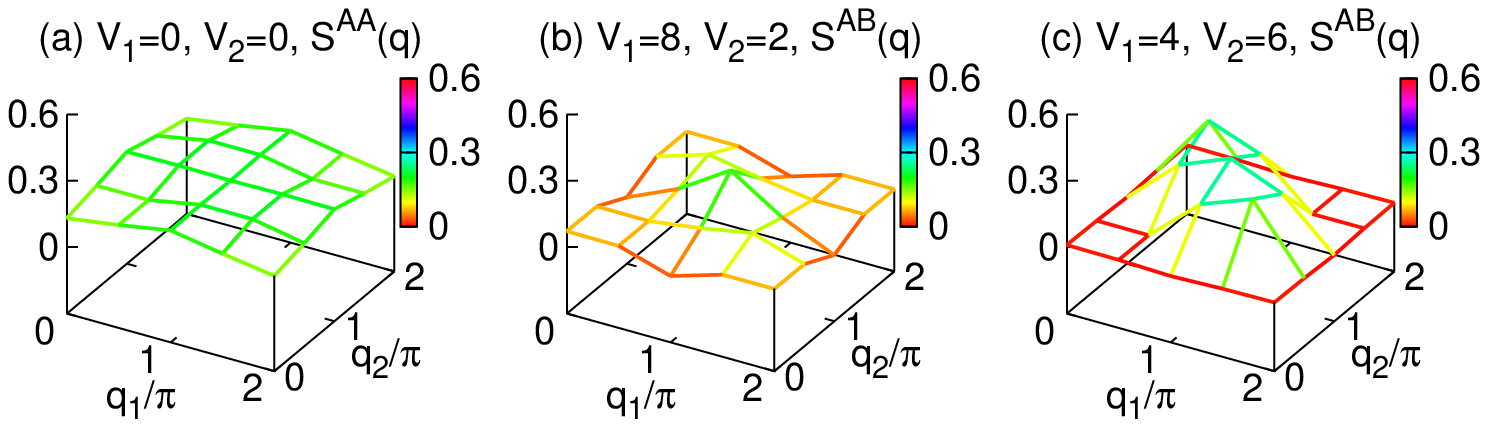}
\par\vspace{0pt}
\end{minipage}
\vspace{-0.12in}
\begin{center}
\begin{minipage}[c]{0.16\textwidth}
\centering
  \includegraphics[scale=0.4]{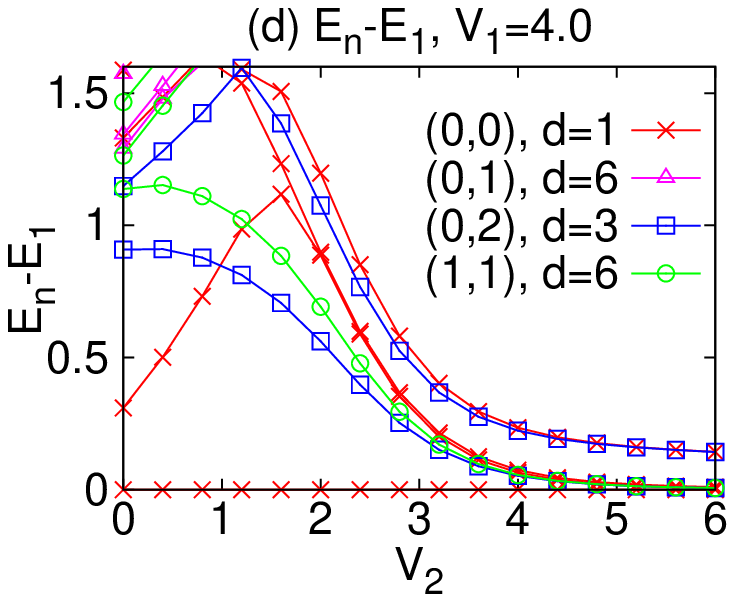}
\par\vspace{0pt}
\end{minipage}%
\hspace{0.008\textwidth}%
\begin{minipage}[c]{0.16\textwidth}
  \includegraphics[scale=0.4]{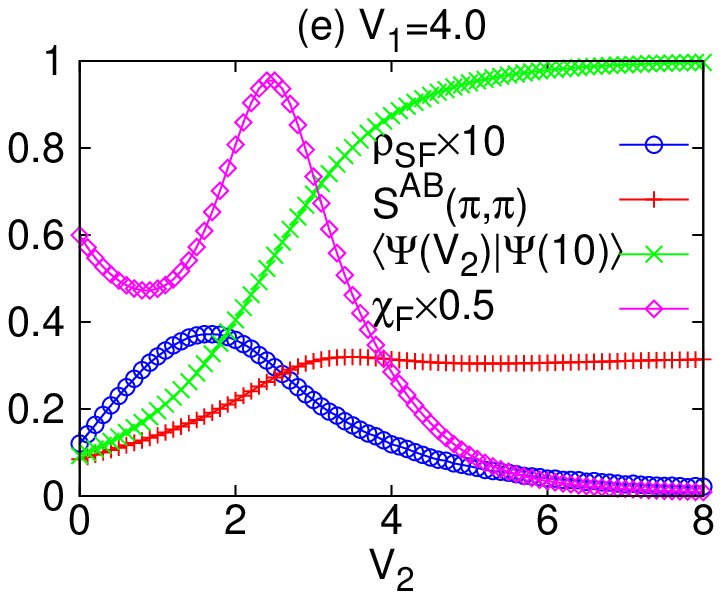}
\par\vspace{0pt}
\end{minipage}
\hspace{0.01\textwidth}%
\begin{minipage}[c]{0.13\textwidth}
  \includegraphics[scale=0.10]{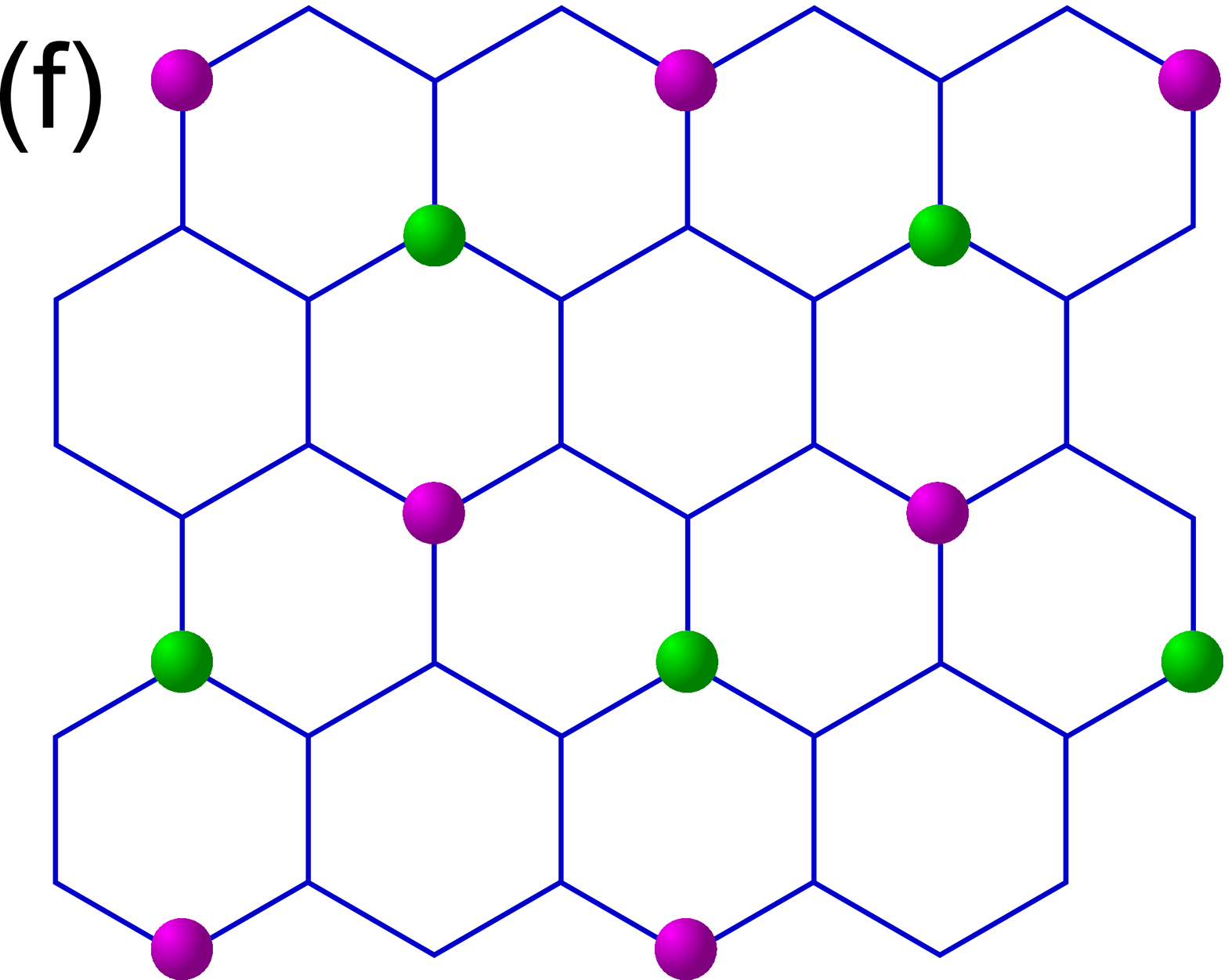}
\par\vspace{-2pt}
\end{minipage}
\end{center}
\vspace{-0.15in}
  \caption{(color online). $32$-site honeycomb lattice at $\nu=1/2$.
   (a) $S^{\rm AA}(\mathbf{q})$ of the FQHE phase;
   (b) $S^{\rm AB}(\mathbf{q})$ of the SF phase;
   (c) $S^{\rm AB}(\mathbf{q})$ of the solid phase.
   (d) Excited energy $E_n-E_1$ in various sectors (with $d$-fold degeneracy) versus $V_2$ at $V_1=4.0$.
   (e) $\rho_{\rm SF}$, $S^{\rm AB}(\pi,\pi)$,
   GS overlap $|\langle\Psi(V_2)|\Psi(10)\rangle|$,
   and fidelity susceptibility $\chi_{\rm F}$ versus $V_2$, at a fixed $V_1=4.0$.
   (f) Illustration of boson occupancy in the solid phase.} \label{f.4}
\end{figure}

{\it SF stiffness and structure factors.---}The $1/2$ FQHE phase on
the honeycomb lattice is also distinguished from the other phases by
the featureless intra-sublattice (AA) structure factor $S^{\rm
AA}(\mathbf{q})$ [Fig.~\ref{f.4}(a)]. The solid phase at a larger
$V_2$ is characterized by a ridge  with $q_1+q_2=2\pi$ in the
inter-sublattice (AB) structure factor $S^{\rm AB}(\mathbf{q})$
[Fig.~\ref{f.4} (c)] and an almost vanishing $\rho_{\rm SF}$
[Fig.~\ref{f.4}(e)]. The SF phase at a smaller $V_2$ has the finite
$\rho_{\rm SF}$ [Fig.~\ref{f.4}(e)] but with a weaker ridge in
$S^{\rm AB}(\mathbf{q})$ [Fig.~\ref{f.4} (b)]. At a fixed $V_1=4.0$
while tuning $V_2$, a transition from the FQHE to the SF phase
occurs with the level crossing of $E_2$ and $E_3$ around $V_2=1.0$
[Fig.~\ref{f.4}(d)]; and a transition from the SF phase to the solid
phase near $V_2=2.5$ is indicated by a peak of the GS fidelity
susceptibility~\cite{SJGu} in Fig.~\ref{f.4}(e), where $\chi_{\rm
F}=2[1-|\langle\Psi(V_2)|\Psi(V_2+\delta)\rangle|]/\delta^2$.

\begin{figure}[!htb]
  \vspace{-0.15in}
  \hspace{0.0in}
\begin{minipage}[b]{0.5\textwidth}
\centering
  \includegraphics[scale=0.55]{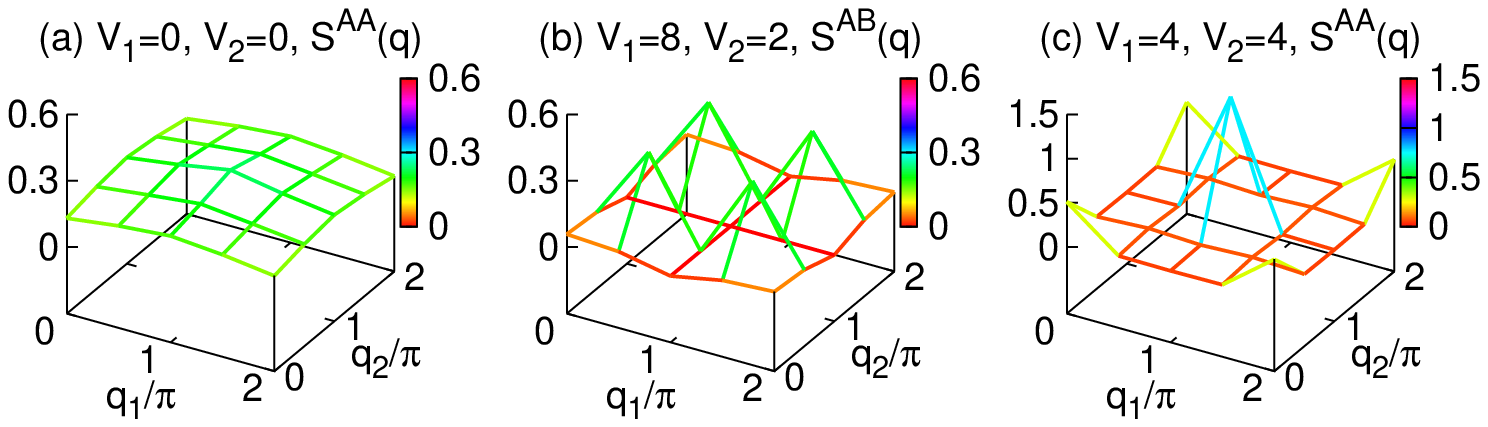}
\par\vspace{0pt}
\end{minipage}
\vspace{-0.15in}
\begin{center}
\begin{minipage}[c]{0.17\textwidth}
  \includegraphics[scale=0.45]{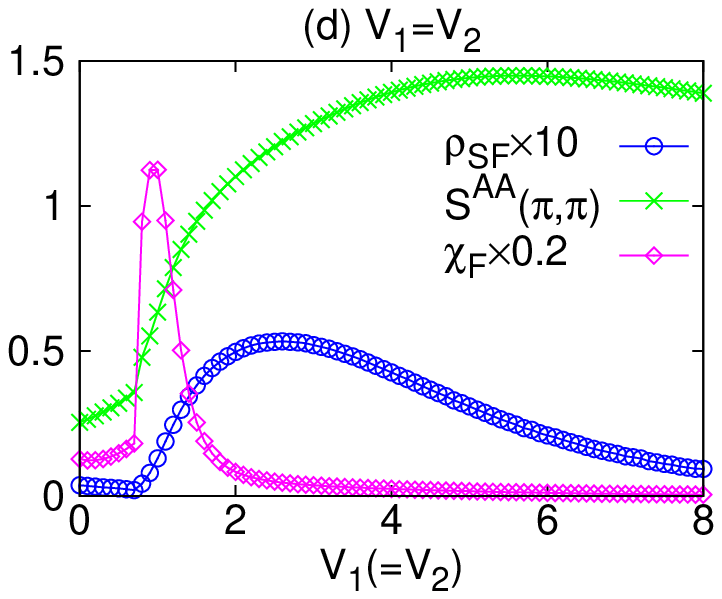}
\par\vspace{0pt}
\end{minipage}%
\hspace{0.015\textwidth}%
\begin{minipage}[c]{0.15\textwidth}
  \includegraphics[scale=0.10]{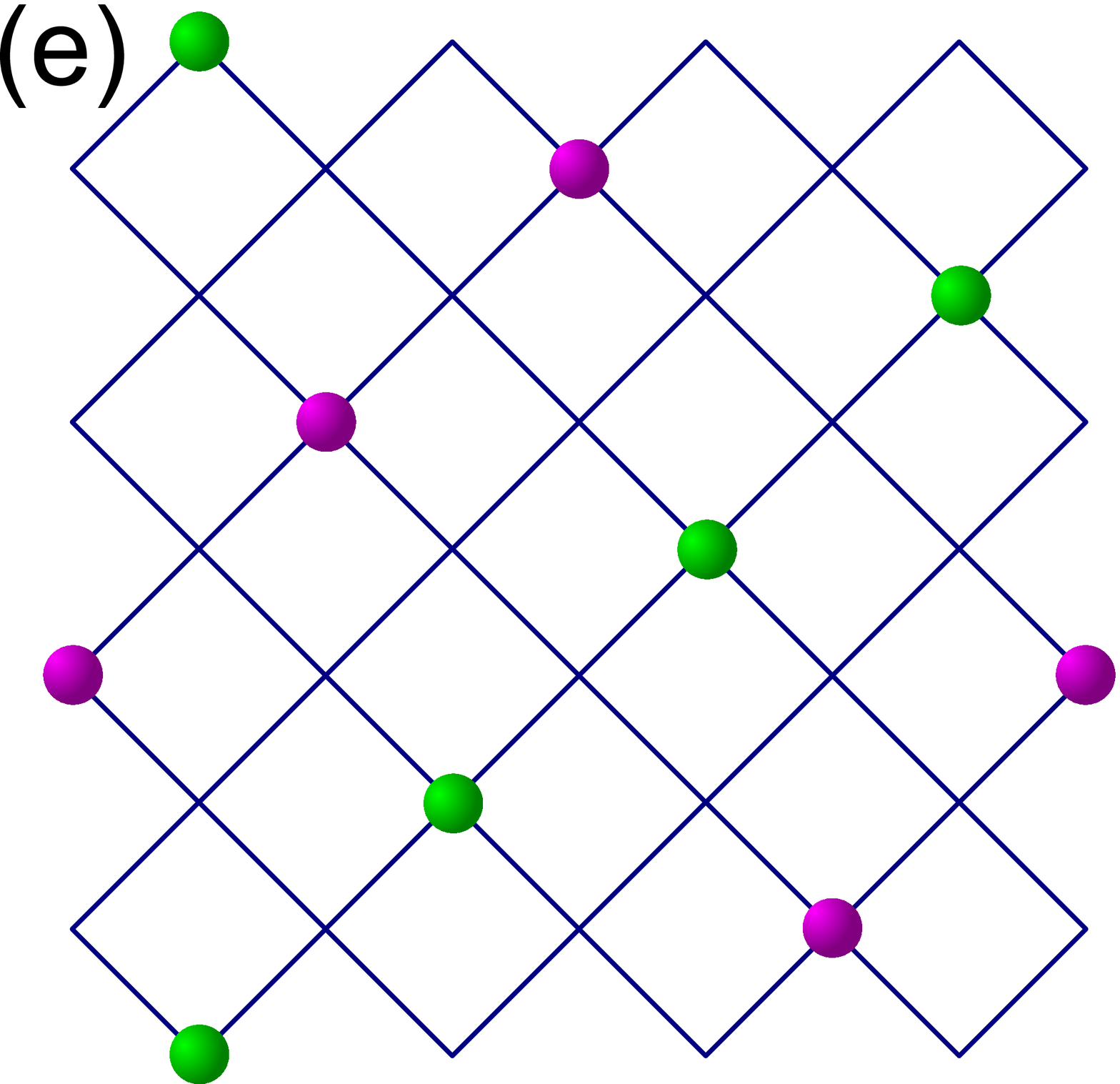}
\par\vspace{0pt}
\end{minipage}
\hspace{-0.02\textwidth}%
\begin{minipage}[c]{0.15\textwidth}
  \includegraphics[scale=0.10]{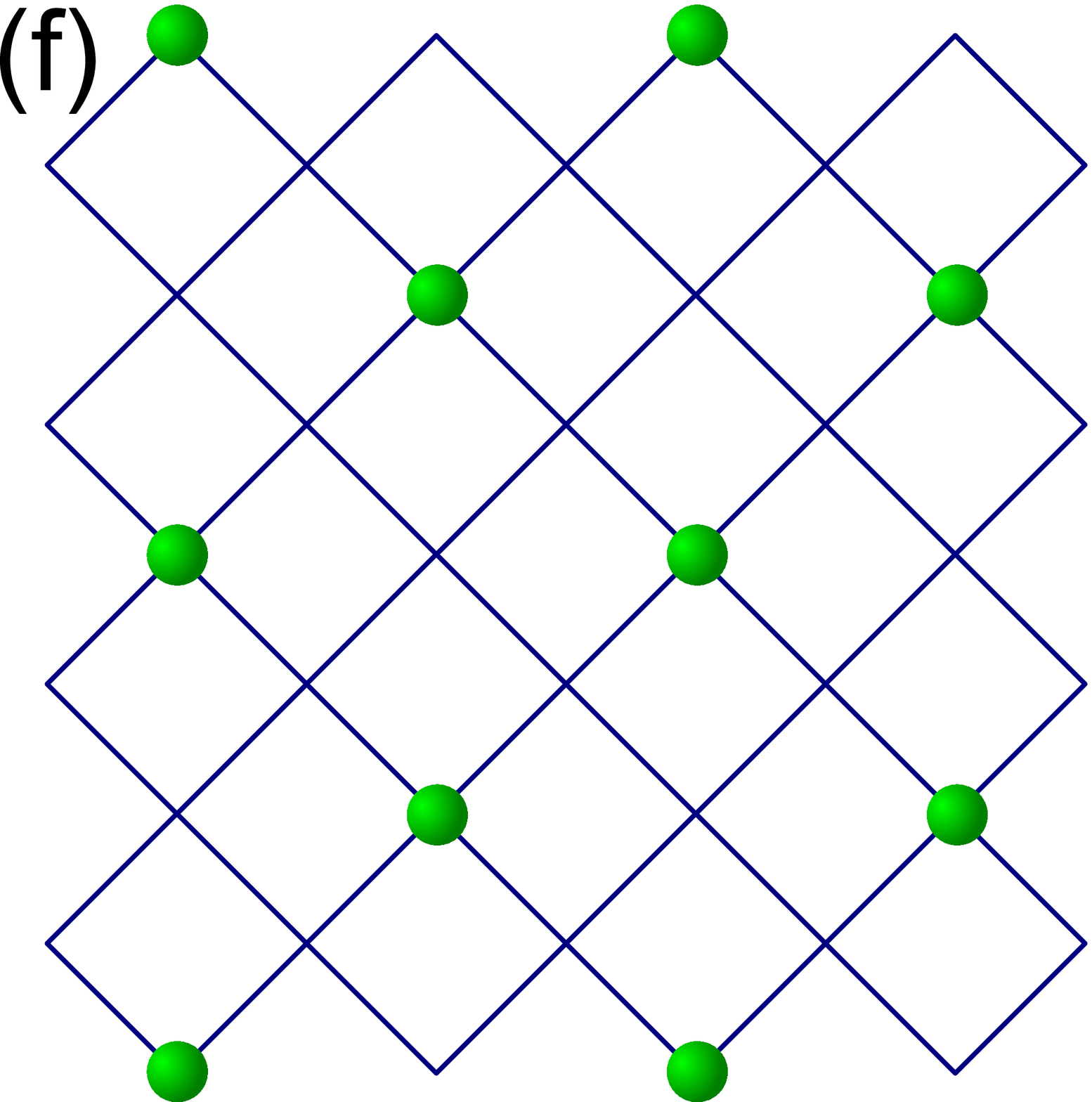}
\par\vspace{0pt}
\end{minipage}
\end{center}
  \vspace{-0.15in}
  \caption{(color online). $32$-site checkerboard lattice at $\nu=1/2$.
   (a) $S^{\rm AA}(\mathbf{q})$ of the FQHE phase.
   (b) $S^{\rm AB}(\mathbf{q})$ of the SS1 phase;
   (c) $S^{\rm AA}(\mathbf{q})$ of the SS2 phase.
   (d) $\rho_{\rm SF}$, $S^{\rm AA}(\pi,\pi)$, and $\chi_{\rm F}$
       versus $V_1$ along the $V_1=V_2$ line.
   (e)-(f) Illustration of boson occupancy in the SS1 and SS2 phases, respectively.} \label{f.5}
\end{figure}

Similarly, the $1/2$ FQHE phase on the checkerboard lattice also
differs from the other phases by featureless $S(\mathbf{q})$
[Fig.~\ref{f.5}(a)]. While both SS1 and SS2 phases are characterized
by either the $\mathbf{q}=(\pi/2,\pi/2)$ peak of $S^{\rm
AB}(\mathbf{q})$ [Fig.~\ref{f.5}(b)] or the $\mathbf{q}=(\pi,\pi)$
peak of $S^{\rm AA}(\mathbf{q})$ [Fig.~\ref{f.5}(c)]. Along the
$V_1=V_2$ line while $V_1(=V_2)$ being tuned, a transition can be
inferred from the FQHE phase to the SS2 phase around $V_1(=V_2)=1.0$
by a sharp peak in the GS fidelity susceptibility
[Fig.~\ref{f.5}(d)]. We emphasize that the firm establishment of the
supersolid and solid phases needs scaling of both $\rho_{\rm SF}$
and $S(\mathbf{q})$ for systems with larger sizes, e.g.
$2\times6\times6$ and $2\times8\times8$, which are compatible with
the ordering patterns
[Figs.~\ref{f.4}(f),~\ref{f.5}(e),~\ref{f.5}(f)] but are far beyond
the capability of the present ED method.

\begin{figure}[!htb]
  \vspace{0.0in}
  \hspace{0.0in}
  \includegraphics[scale=0.50]{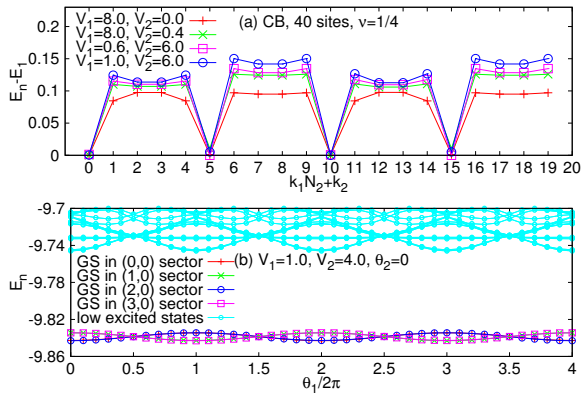}
  \vspace{-0.06in}
  \caption{(color online). The $1/4$-FQHE of the $40$-site checkerboard
  model: (a) spectrum gaps $E_n-E_1$ versus $k_1N_2+k_2$;
  (b) low energy spectra versus $\theta_1$ at a fixed $\theta_2=0$. } \label{f.6}
\end{figure}

{\it The $\nu=1/4$ FQHE.---}We searched for the $\nu=1/4$ FQHE for
both lattice models with the system sizes of $N_s=24$, $32$, $40$
and $48$. In contrast to the $\nu=1/2$ case, the onset of the
$1/4$-FQHE needs finite values of either $V_1$ or $V_2$. Examples of
the $40$-site checkerboard system are chosen to demonstrate the
basic properties of the $1/4$-FQHE. For each set of $V_1$ and $V_2$
in Fig.~\ref{f.6}(a), there is clearly a GSM with four states. For
each GSM, the four states evolve into each other when tuning the
boundary phases [Fig.~\ref{f.6}(b)], and all the four states share a
total Chern number $1$.  These are concrete evidences of the
$1/4$-FQHE in some small parameter regions, however, these regions
depend on the lattice sizes more sensitively than those of the
$1/2$-FQHE. We conjecture that a finite NNNN repulsion $V_3$ may be
necessary to get a large and stable parameter space of the
$1/4$-FQHE, which will be addressed in a future work.

{\it Summary and discussion.---}We consider hard-core bosons in two
representative TFB models with NN and NNN repulsions. We find
convincing numerical evidences of both the $1/2$ and the $1/4$
bosonic FQHE phases which are characterized by distinctive finite
spectrum gap; quasi-degenerate states in a GSM which can evolve into
each other upon varying boundary phases; smooth Berry curvature and
topologically invariant unit total Chern number for the GSM. For
both lattices, the $1/2$-FQHE phase is found to occupy a significant
space of phase diagrams, in addition to other conventional ordered
phases. Interestingly, such a $1/2$-FQHE is very stable (large
spectrum gap) for hard-core bosons even without additional
interactions ($V_1=V_2=0$), which makes it easier to be realized by
cold atoms in optical lattices.

Our work here focuses on hard-core bosons in TFB models, while a
recent parallel work found the $1/3$ and $1/5$ FQHE of interacting
fermions on the checkerboard model~\cite{Sheng2} (see also an
updated version of Ref.~\cite{Santos}). There is also an interesting
proposal lately to find the bosonic FQHE (in terms of chiral spin
states) in frustrated kagom\'{e}-lattice magnets~\cite{WenCSS}.

This work is supported by DOE Office of Basic Energy Sciences under
grant DE-FG02-06ER46305 (DNS). We also acknowledge the NSFC of China
grant No. 10904130 (YFW), the US NSF grant No. NSFPHY05-51164 (ZCG),
and the State Key Program for Basic Researches of China grants No.
2006CB921802 and No. 2009CB929504 (CDG).

\end{document}